\documentclass{emulateapj}

\slugcomment{Submitted to the Astrophysical Journal }

\shorttitle{Plasma-wave generation in a dynamic spacetime}
\shortauthors{Huan Yang and Fan Zhang}

\usepackage{amsmath,amssymb}
\usepackage[normalem]{ulem}
\usepackage{bm}
\usepackage{comment}
\usepackage{xcolor}
\usepackage{soul}
\usepackage{hyperref}
\usepackage[percent]{overpic}
\usepackage{overpic}
\usepackage{natbib}

\begin{document}

\newcommand{\bea}{\begin{eqnarray}}
\newcommand{\eea}{\end{eqnarray}}
\newcommand{\E}{\mathrm{E}}
\newcommand{\Var}{\mathrm{Var}}
\newcommand{\bra}[1]{\langle #1|}
\newcommand{\ket}[1]{|#1\rangle}
\newcommand{\braket}[2]{\langle #1|#2 \rangle}
\newcommand{\mean}[2]{\langle #1 #2 \rangle}
\newcommand{\be}{\begin{equation}}
\newcommand{\ee}{\end{equation}}	
\newcommand{\ba}{\begin{eqnarray}}
\newcommand{\ea}{\end{eqnarray}}
\newcommand{\SD}[1]{{\color{magenta}#1}}
\newcommand{\rem}[1]{{\sout{#1}}}
\newcommand{\alert}[1]{\textbf{\color{red} \uwave{#1}}}
\newcommand{\Y}[1]{\textcolor{blue}{#1}}
\newcommand{\R}[1]{\textcolor{red}{#1}}
\newcommand{\B}[1]{\textcolor{black}{#1}}
\newcommand{\C}[1]{\textcolor{cyan}{#1}}
\newcommand{\db}{\color{darkblue}}
\newcommand{\huan}[1]{\textcolor{cyan}{#1}}
\newcommand{\fan}[1]{\textcolor{blue}{#1}}
\newcommand{\ac}[1]{\textcolor{cyan}{\sout{#1}}}
\newcommand{\intinfty}{\int_{-\infty}^{\infty}\!}
\newcommand{\Tr}{\mathop{\rm Tr}\nolimits}
\newcommand{\const}{\mathop{\rm const}\nolimits}

\title{Plasma-wave generation in a dynamic spacetime}

\author{Huan Yang        \altaffilmark{1,2} and
        Fan Zhang  \altaffilmark{3,4},
}

\altaffiltext{1}{Perimeter Institute for Theoretical Physics, Waterloo, Ontario N2L2Y5, Canada}
\altaffiltext{2}{Institute for Quantum Computing, University of Waterloo, Waterloo, Ontario N2L3G1, Canada}
\altaffiltext{3}{Gravitational Wave and Cosmology Laboratory, Department of Astronomy, Beijing Normal University, Beijing 100875, China}
\altaffiltext{4}{Department of Physics and Astronomy, West Virginia University, PO Box 6315, Morgantown, WV 26506, USA}

\begin{abstract}
We propose a new electromagnetic-emission mechanism in  magnetized, force-free plasma, which is driven by the evolution of the underlying dynamic spacetime. 
In particular, the emission power and angular distribution of the emitted fast-magnetosonic and Alfv\'en waves are separately determined. Previous numerical simulations of binary black hole mergers occurring within magnetized plasma have recorded copious amounts of electromagnetic radiation that, in addition to collimated jets, include an unexplained,  isotropic component which becomes dominant close to merger. This raises the possibility of multimessenger gravitational-wave and electromagnetic observations on binary black hole systems. The mechanism proposed here provides a candidate analytical characterization of the numerical results, and when combined with previously understood mechanisms such as the Blandford-Znajek process and kinetic-motion-driven radiation, allows us to construct a classification of different electromagnetic radiation components seen in the inspiral stage of compact-binary coalescences. 
\end{abstract}


\section{Introduction}
With the imminent direct detection of gravitational waves (GWs) by second generation detectors (\cite{Dooley:2014iga}),
the pursuit of an understanding of the electromagnetic (EM) counterparts to GWs becomes urgent, as a joint observation in both channels will provide irreplaceable means to diagnose properties of the astrophysical sources (\cite{Christensen:2011qf}). One of the most important types of sources that could radiate both gravitationally and electromagnetically is a coalescing compact binary, involving black holes and/or neutron stars surrounded by magnetized plasma (forming the so-called ``magnetospheres''). The magnetic field could originate from the accretion disk of the binary or neutron stars themselves, and the plasma could be generated from vacuum polarization, and/or charged particles coming off of the star surfaces and the accretion disk. Recent numerical simulations (\cite{Palenzuela:2010nf, Neilsen:2010ax, Alic:2012df}) have shown that EM radiation is indeed given off by such systems in abundance even \emph{before merger} and for \emph{binary black hole} systems (while current joint-observation efforts concentrate on the post-merger stage of systems with at least one neutron star (\cite{Nissanke:2012dj})), providing further optimism for the success of multi-messenger astronomy. The next step is then to clarify the various physical processes at work that, together, produces the EM signals seen numerically (in particular, an isotropic radiation that dominates near merger time has not been previously understood analytically). A complete classification and characterization of these processes is a prerequisite for extracting useful information about the binary systems from the observed EM signals. We provide such an analytical characterization in this work and compare it with previous numerical results (see Fig.~\ref{fig:comparison} below).

Within magnetospheres, the energy density of the magnetic field often dominates over that of the plasma particles, creating what's referred to as a {\it force-free plasma}. 
Thanks to the seminal works by \cite{Goldreich:1969sb} and \cite{1977MNRAS.179..433B}, it is widely accepted that force-free plasma can act as a medium for powering outgoing EM radiation (or jets) at a cost of reducing the rotational energy of neutrons stars or black holes (\cite{thorne:BlackHolesAndTimeWarps,Meier:1481607,2011CQGra..28m4007P,1997MNRAS.288..333S,Hansen:2000am}). More recent studies (\cite{Hansen:2000am,Lyutikov:2011vca,Brennan:2013ppa, Penna:2015qta, Palenzuela:2009yr, Palenzuela:2009hx, Palenzuela:2010nf, Palenzuela:2010xn, Neilsen:2010ax, Moesta:2011bn, Alic:2012df, Paschalidis:2013jsa,McWilliams:2011zi,D'Orazio:2013kgo,Morozova:2013ina}) suggest that a force-free plasma could also drain the (linear-motion) kinetic energy of moving objects to power EM radiations in the form of jets launching from star surfaces (or the black hole horizon), accompanied by some isotropic flux. We refer to this as the {\it kinetic-motion-driven radiation} \footnote{We caution that although terms like this have been used, here and in other literature, to label different EM emission mechanisms, the nonlinearity of force-free dynamics and gravity makes a mathematically rigorous classification difficult. This is particularly true with fully nonlinear numerical simulations, in which all of the emission mechanisms discussed here and below are likely present, even when the simulation aims to study a particular one.}, which is also seen from satellites moving in earth's ionosphere (\cite{Drell1965, Drell1965prl}).

There is however, a third mechanism, which we shall call the {\it gravitation-driven radiation}, which will be the focus of this paper. When the background spacetime becomes dynamic, the local EM energy density of magnetized plasma deviates from its equilibrium values and these inhomogeneities tend to propagate out via plasma waves. A similar phenomenon is known to exist in spacetimes without matter (the Gertsenshtein-Zeldovich effect (\cite{Gertsenshtein,Zeldovich1973})), where the outgoing radiation consists purely of vacuum EM waves. 
In addition, the generation of magnetohydrodynamic (MHD) waves by the influence of gravitational waves has been examined in \cite{Duez:2005sg}.  
Although this effect has not been explicitly discussed in the context of force-free magnetospheres, we note that force-free electrodynamics (FFE) can be viewed as the low-inertia limit of relativistic magnetohydrodynamics (\cite{McKinney:2006sc,Paschalidis:2013gma}). 
In this paper, we will examine essentially the same physical process, but where the driving gravitational dynamics is not an (idealized wave-zone) gravitational wave.
Within force-free plasma, energy can be carried away by two different classes of waves. One class is called the fast-magnetosonic waves in the local short-wavelength limit (the wavelength is much smaller than the radius of spacetime curvature), whose global and longer-wavelength counterparts are named the ``trapped modes'' in \cite{Yang:2014zva,Yang:2015ata}. These tend to behave similarly to vacuum EM waves and propagate in a more egalitarian fashion in terms of sky directions. 
The other class of waves are the Alfv\'en waves, generalizing to ``traveling waves" (\cite{Yang:2014zva,Yang:2015ata}) or principal null solutions (\cite{Brennan:2013jla,Zhang:2015aga}). A salient feature of the Alfv\'en waves and their generalizations (for brevity, we will not distinguish between them below, similarly for the other class) is that they propagate along the magnetic field lines, and are as such automatically collimated if the magnetic field threads through the orbital plane of the binary nearly orthogonally (a natural configuration for accretion-disk-supported field). Below, we show how to compute their fluxes as generated by the gravitationally-driven process. 

In order to perform the analysis, we apply the geometric approach promoted by \cite{Gralla:2014yja,1997MNRAS.286..931U,1998MNRAS.297..315U,
1997MNRAS.291..125U,Carter1979,1997PhRvE..56.2181U,1997PhRvE..56.2198U}, whose introduction has triggered many new developments (\cite{Zhang:2014pla, Gralla:2015wva, Lupsasca:2014pfa, Lupsasca:2014hua,Gralla:2015vta,Gralla:2015uta}) in obtaining exact solutions to force-free electrodynamics, in addition to new interpretations of previous results (\cite{Penna:2014aza, Menon2005}). Unless otherwise specified, the formulae below are in natural units, so that $c=G=1$.

\section{Set-up of the problem}
Let us assume that there is a stationary FFE configuration in a stationary background spacetime with metric $g_{\rm B}$. According to discussions in \cite{1997PhRvE..56.2181U,1997PhRvE..56.2198U,Gralla:2014yja}, it is possible to find at least one pair of ``Euler potentials" $\phi_{1B,2B}$, such that
$F_{\rm B} = d \phi_{\rm 1B} \wedge d \phi_{\rm 2B}$,
where $F_{\rm B}$ is the background electromagnetic field tensor. Now suppose that the spacetime becomes dynamic and its metric is $g =g_{\rm B}+\epsilon h$, where $\epsilon$ parametrizes the magnitude of the spacetime deformation from its stationary state. Correspondingly the Euler potentials will also deviate from their original values: $\phi_{1,2} = \phi_{\rm 1,2 B}+\epsilon \delta \phi_{1,2}$, whereby the non-linear FFE wave equations they satisfy are \cite{Gralla:2014yja}
\begin{align} \label{eq:FFEEq}
d \phi_{1,2} \wedge d * F=0\,,
\end{align} 
with $F \equiv d \phi_1 \wedge d \phi_2$. Note that the Hodge star $*$ is now with respect to the total metric $g$, so that it depends on metric perturbations. In order to study the gravitationally-induced plasma waves, we shall linearise the above equation to the leading order in $\epsilon$, and obtain
\begin{align}\label{eqgen}
d \delta \phi_{1,2} \wedge d *_{\rm B} F_{\rm B}+& d\phi_{\rm 1,2 B} \wedge d *_{\rm B} \delta F \nonumber \\
 =& - d \phi_{\rm 1,2 B} \wedge d \frac{\partial * F_{\rm B}}{\partial \epsilon} \,.
\end{align}
This equation describes the excitation of plasma fields $\delta \phi_{1,2}$ by the source on the right hand side, which is linear in $h$. It implies that GWs interacting with magnetized plasma can generate plasma waves. Moreover, it  predicts that a time-dependent Newtonian source within magnetized plasma also induces plasma  radiation, an effect that has been overlooked before and could have observational consequences.

\section{Radiation in nearly flat spacetimes}
Now we specialize to a simple yet important example where the background metric is flat, i.e., $g_{ \mu\nu}=\eta_{\mu\nu}+\epsilon h_{\mu\nu}$. This is a good approximation when the gravitational field generated by matter sources or GWs is weak. In addition, let us assume that the plasma is magnetized along the $z$ direction, with field strength $B$ so that $F_{\rm B} =B dx \wedge dy$. When the spacetime becomes dynamic, the EM field $2$-form can be written as (note we consider only those FFE perturbations driven by the spacetime variations, and so use the same flag $\epsilon$)
\begin{align} \label{eq:Faraday}
F = B (dx+\epsilon \delta \phi_1) \wedge (dy +\epsilon \delta \phi_2)\,.
\end{align}
With this set-up, one can straightforwardly work out the Hodge star rules, plug them into Eq.~\eqref{eqgen}, and obtain a coupled set of wave equations for $\delta \phi_{1,2}$. These equations can further be diagonalized through the definition of a new set of variables:
\begin{align}\label{eqdef}
\psi_1 \equiv \partial_x \delta \phi_2-\partial_y \delta \phi_1,\quad \psi_2 \equiv\partial_y \delta \phi_2+\partial_x \delta \phi_1\,,
\end{align}
in which case the wave equations decouple into 
\begin{align}\label{eqwe}
 (-\partial^2_t+\partial^2_z)\psi_1 =& \frac{\partial^2 h_{tx}}{\partial t \partial y}-\frac{\partial^2 h_{ty}}{\partial t \partial x}+\frac{\partial^2 h_{yz}}{\partial z \partial x}-\frac{\partial^2 h_{xz}}{\partial z \partial y}\,, \nonumber \\
 (-\partial^2_t+\nabla^2) \psi_2 =&\frac{1}{2} (\partial^2_x+\partial^2_y)(h_{tt}+h_{xx}+h_{yy}-h_{zz}) \nonumber \\+\frac{\partial^2 h_{yz}}{\partial y \partial z} 
&+\frac{\partial^2 h_{xz}}{\partial x \partial z}-\frac{\partial^2 h_{yt}}{\partial y \partial t}-\frac{\partial^2 h_{xt}}{\partial x \partial t}\,.
\end{align}
The first equation describes a wave propagating along the magnetic field lines, or in other words the Alfv\'en wave.  The second equation describes the fast-magnetosonic wave, which propagates in all directions. These equations are gauge-invariant, as can be checked by substituting in the infinitesimal gauge transformation $x_i \rightarrow x_i+\xi_i$ that leads to 
\bea
h_{\mu\nu} \rightarrow h_{\mu\nu}+\xi_{\mu|\nu}+\xi_{\nu | \mu} \approx h_{\mu\nu}+\xi_{\mu,\nu}+\xi_{\nu,\mu}\,,
\eea
and
\begin{equation}
\delta \phi_1 \rightarrow \delta \phi_1 + \xi^x,\quad \delta \phi_2\rightarrow \delta \phi_2 +\xi^y\,.
\end{equation}
 Denote the source terms in Eq.~\ref{eqwe} as $S_{1}$ and $S_{2}$ respectively, the solutions to these wave equations can be obtained through the use of Green's functions, 
\begin{align}\label{eqreds}
\psi_1 &=\frac{1}{2} \int dz dt \,\Theta(t-t'-|z-z'|) S_1(t',z')\,,\nonumber \\
\psi_2
& = -\int d^3{\bf x}\frac{S_2(t-|{\bf x}-{\bf x'}|,{\bf x'})}{4 \pi |{\bf x}-{\bf x'}|}\,,
\end{align} 
where $\Theta$ denotes the Heaviside step function. After evaluating $\psi_{1,2}$, we can reconstruct $\delta \phi_{1,2}$, and subsequently $F$, by noting that
\begin{align}
&(\partial^2_x+\partial^2_y)\delta \phi_2 = \partial_x \psi_1+\partial_y \psi_2\,,\nonumber \\
&(\partial^2_x+\partial^2_y)\delta \phi_1 = -\partial_y \psi_1+\partial_x \psi_2\,,
\end{align}
whose solutions are (applying the Green's function for 2-D elliptic equations)
\begin{align}\label{eqrecon}
&\delta \phi_2 = \int dx' dy' \frac{\Delta_x\psi_1(x',y',z)+\Delta_y\psi_2(x',y',z)}{2 \pi (\Delta_x^2+\Delta_y^2)}\,,\nonumber \\
&\delta \phi_1 = \int dx' dy' \frac{\Delta_x\psi_2(x',y',z)-\Delta_y\psi_1(x',y',z)}{2 \pi (\Delta_x^2+\Delta_y^2)}\,,
\end{align}
where $\Delta_x= x-x'$ and $\Delta_y=y-y'$.

Analogous to the Gertsenshtein-Zeldovich effect, Eq.~\eqref{eqwe} together with Eq.~\eqref{eqreds} explicitly show that GWs injected into magnetized plasma would generate both Alfv\'en and fast-magnetosonic waves. Suppose that the gravitational wave packet has a characteristic amplitude $h$ and a length-scale of $\lambda$, it is then straightforward to see that the plasma-wave luminosity $\mathcal{L_{\rm GW}}$ satisfies $\mathcal{L_{\rm GW}}\propto B^2 \lambda^2 h^2$; a relationship that can be compared with future numerical experiments. Here we focus instead on the case where the source is generated by two orbiting compact masses, in order to study the radiation of a binary system in the inspiral stage. With a Newtonian matter source (as the leading order post-Newtonian term of general relativistic expressions, which is sufficient for our purpose), $h$ is given by (\cite{MTW})
\begin{align}\label{eqh}
h_{00} & = 2 \int d^3 x'\frac{\rho(x')}{|{\bf x}-{\bf x'}|} \,, \quad 
h_{0j} & =-4 \int d^3 x'\frac{\rho(x') v_j'}{|{\bf x}-{\bf x'}|} \,,\nonumber \\
h_{jk} & = 2 \delta_{jk} \int d^3 x'\frac{\rho(x')}{|{\bf x}-{\bf x'}|}\,.
\end{align}
When the source consists of a pair of orbiting black holes, the formulae above are valid at places away from the black holes, which are themselves replaced by point masses. 
However, the Newtonian approximation becomes inaccurate near the black holes. In addition, in order to compute the plasma waves at far away and extract the energy flux, we must exclude the points enclosed by the black hole horizons. Therefore, in practise (see Sec.~\ref{sec5}), we remove two excision spheres when computing the integrals in Eq.~\ref{eqreds}. To test the sensitivity of the gravitation-driven luminosity values on the excision radii choice, we vary their values from $2\tilde{M}$ to $3\tilde{M}$ ($\tilde{M}$ being the black-hole mass), and observed that the resulting flux changes less than $10$ precent. For the presentation of data in Sec.~\ref{sec5} then, we adopt the cut-off radius choice of $3\tilde{M}$. We caution that this insensitivity to excision radii could change significantly if we take into account relativistic (Post-Newtonian) corrections to the metric.

\section{Flux extraction}
According to Eq.~\eqref{eqreds}, the fast-magnetosonic waves are quite similar to the vacuum EM waves, where the source term $S_2$ can also be decomposed into multipole contributions. Let us assume that the binary (with total mass M) is practicing near-circular motion with a period of $2\pi/\Omega$, in which case $\psi_2$ in the radiative zone can be written as
\begin{align}\label{eqasym2}
\psi_2 \sim \sum_m f_m(\theta) \frac{e^{i m [\phi -\Omega(t-r)]}}{r}\,,
\end{align}
where the $m=0$ piece corresponds to the DC monopole field, which does not radiate. The coefficients $f_m$ may be further decomposed into a summation of associated Lengendre polynomials, starting from $l \ge |m|$. In order to compute the energy flux, we need to reconstruct $\delta \phi_{1,2}$ with Eq.~\eqref{eqrecon} (in the absence of $\psi_1$), or more efficiently, by noticing that $\delta \phi_{1,2}$ must possess similar asymptotic forms as Eq.~\eqref{eqasym2}:
\begin{align}\label{eqasymphi}
\delta \phi_{1,2} \sim \sum_m g^{1,2}_m(\theta) \frac{e^{i m [\phi -\Omega(t-r)]}}{r}\,,
\end{align}
and the relationship between $g^{1,2}_m$ and $f_m$ can be obtained using Eq.~\eqref{eqdef} with $\psi_1=0$:
\begin{align}
g^{(1)}_m(\theta) =& -\frac{i }{2 m \Omega \sin \theta} [f_{m+1}(\theta)+f_{m-1}(\theta)]\,, \notag \\
g^{(2)}_m(\theta) =& -\frac{1}{2 m \Omega \sin \theta} [f_{m+1}(\theta)-f_{m-1}(\theta)]\,.
\end{align}
We can then substitute these expressions into $\delta\phi_{1}$ and $\delta\phi_{2}$, and subsequently Eq.~\eqref{eq:Faraday} to obtain the field 2-form. It is then straightforward, although tedious, to extract from it the electric and magnetic field vectors, and compute the Poynting vector. In the end, we arrive at the flux formula for fast-magnetosonic waves:
\begin{align}\label{eqfm}
{S}_{\rm fast}=\sum_{m \neq 0}\frac{B^2 |f_{m}(\theta )|^2 \csc ^2(\theta )}{  r^2}\,.
\end{align}

The Alfv\'en waves, on the other hand, propagate along the magnetic field lines. Based on Eq.~\eqref{eqreds}, we write $\psi_1$ in the radiative zone $|z| \gg M$ as
\bea\label{eqfourier}
\psi_1 
= \int d k_x d k_y A^\pm(k_x,k_y,u,v)e^{i k_x x+i k_y y}\,.
\eea
where $\pm$ stands for the top/down extraction surfaces and $u \equiv t-z, \,\, v \equiv t+z$. The effective radiative part of $\psi_1$ is only a function of $u$ for $z \gg M$, and a function of $v$ for $-z \gg M$. 
One can write the associated $\delta \phi_{1,2}$ in a similar format, which satisfies Eq.~\eqref{eqdef} with $\psi_2=0$:
\begin{align}
\delta \phi_2 & =  \int d k_x d k_y \frac{-i k_x}{k^2_x+k^2_y} \mathcal{A}^{\pm}e^{i k_x x+i k_y y}\,, \notag \\
\delta \phi_1  &=  \int d k_x d k_y \frac{i k_y}{k^2_x+k^2_y} \mathcal{A}^{\pm} e^{i k_x x+i k_y y}\,, \notag \\
\mathcal{A}^{\pm} &= A^\pm_v(k_x,k_y,v) +A^\pm_u(k_x, k_y, u)\,,
\end{align}
from which we obtain the luminosity function
\begin{align}\label{eqla}
\mathcal{L}_{\rm Alf} =2 B^2 \sum_{\pm} \int d k_x dk_y \frac{\pm|\partial_u A^\pm|^2\mp|\partial_v A^\pm|^2}{k^2_x+k^2_y}\,.
\end{align}
For systems with mirror symmetry about the orbital plane, it suffices to only compute the luminosity on one side and double the result.

\section{Binary black hole coalescence}\label{sec5}
We can now compare our analytical predictions with numerical simulations of equal-mass binary black hole coalescences, and try to identify the physical mechanisms behind the ``isotropic" and ``collimated" EM radiations seen there (\cite{Palenzuela:2010nf, Neilsen:2010ax, Alic:2012df, Brennan:2013ppa}), as well as to estimate the magnitude of each piece. 
To facilitate comparison, we adopt the same contextual parameters as in the numerical experiments above, i.e., a binary black hole system with $10^8$ solar masses for each hole and a background magnetic field at $10^4$ Gauss. We also note that the strength of the EM emissions is much weaker than that of the gravitational-wave emission, where the gravitational radiation-reaction leads to the shrinking of the orbital radius. As a result, it is a valid and common approximation to ignore any back-reaction of the EM radiations on the evolution of the spacetime.

Both fast-magnetosonic and Alfv\'en waves are produced during the sequence (inspiral, merger, and then ringdown) of binary merger stages, and they radiate \emph{mostly} in the forms of ``isotropic" and ``collimated" fluxes, respectively. 
Below, we will concentrate on the inspiral stage (leading into the merger itself) that's the most interesting for multi-messenger astronomy. 
During this stage, the EM emissions can be classified into rotation-driven, kinetic-motion-driven and gravitation-driven types. The rotation-driven radiation is generated by the Blandford-Znajek mechanism, which supports a jet-like radiation with luminosity of the order (\cite{Neilsen:2010ax})
\begin{align}\label{eqr}
\mathcal{L}_{\rm r} \sim 2.4 \times 10^{43}{\rm  ergs/s} \left ( \frac{B}{10^4 {\rm G}}\right)^2\left (\frac{M_i}{10^8 M_{\odot}} \right)^2 \bar{a}^2_i\,
\end{align}
in cgs units and when spin is aligned with the magnetic field, or abbreviated as $2.4 L_{43}B^2_4 M^2_{i 8}\bar{a}^2_i$. Here $M_i$   is the $i$th black hole mass and $\bar{a}_i$ is the dimensionless spin parameter of the black hole ranging from $0$ to $1$. 

As a black hole moves through magnetized force-free plasma, it launches collimated jets along the magnetic field lines (\cite{Palenzuela:2010nf, Neilsen:2010ax}). The power of this radiation is proportional to $v^2$  and thus $\Omega^{2/3}$. In addition, if the black hole also follows accelerated motion,  it generates an additional Poynting flux similar to accelerated charges in vacuum, which can be attributed to fast-magnetosonic wave emission. Its power is on the order of $2/3 q^2 a^2$  (``Larmor formula'' of  \cite{Brennan:2013ppa}), where the effective monopole charge $q$ should have value $\sim 2 B M^2$ and the acceleration obeys $a \propto v^2/d \propto  \Omega^{4/3}$.
Summing up the two contributions, we have for kinetic-motion-driven radiation that
\begin{align} \label{eq:MotionDriven}
\mathcal{L}_{\rm m} \sim 1.6  L_{43} B^2_4 M^{8/3}_8 \Omega_{-4}^{2/3}+ 0.5  L_{40} B^2_4 M^{14/3}_8 \Omega_{-4}^{8/3}\,.
\end{align}
The merger happens at around $\Omega \sim2 \times 10^{-4} s^{-1}$, and so the acceleration-induced radiation is sub-dominant through the entire inspiral stage.

We now turn to the gravitation-driven radiation. 
With Eqs.~\eqref{eqwe}, \eqref{eqreds}, and \eqref{eqh}, we can estimate the orbital frequency dependence of this class of EM emissions for a binary black hole system. The source term of fast-magnetosonic waves scales as $M/d^3$, where $d$ is the orbital separation. Such a source term generates $\psi_2$ in the multipolar-expansion manner of Eq.~\eqref{eqasym2}, with the luminosity for each multipole moment scaling as $B^2 M^2 v^{2l} \propto \Omega^{2l/3}$. 
For unequal mass binaries, the radiation contains a dipole piece with $l=1$, whereas emission from an equal-mass binary starts at the quadrupolar order ($l=2$). On the other hand,
the source term for Alfv\'en waves scales as $M v \Omega/d^2$ and the corresponding flux scales as $B^2 M^2 v^4 \propto \Omega^{4/3}$.

\begin{figure}[t,b]
\begin{overpic}[width=0.95\columnwidth]{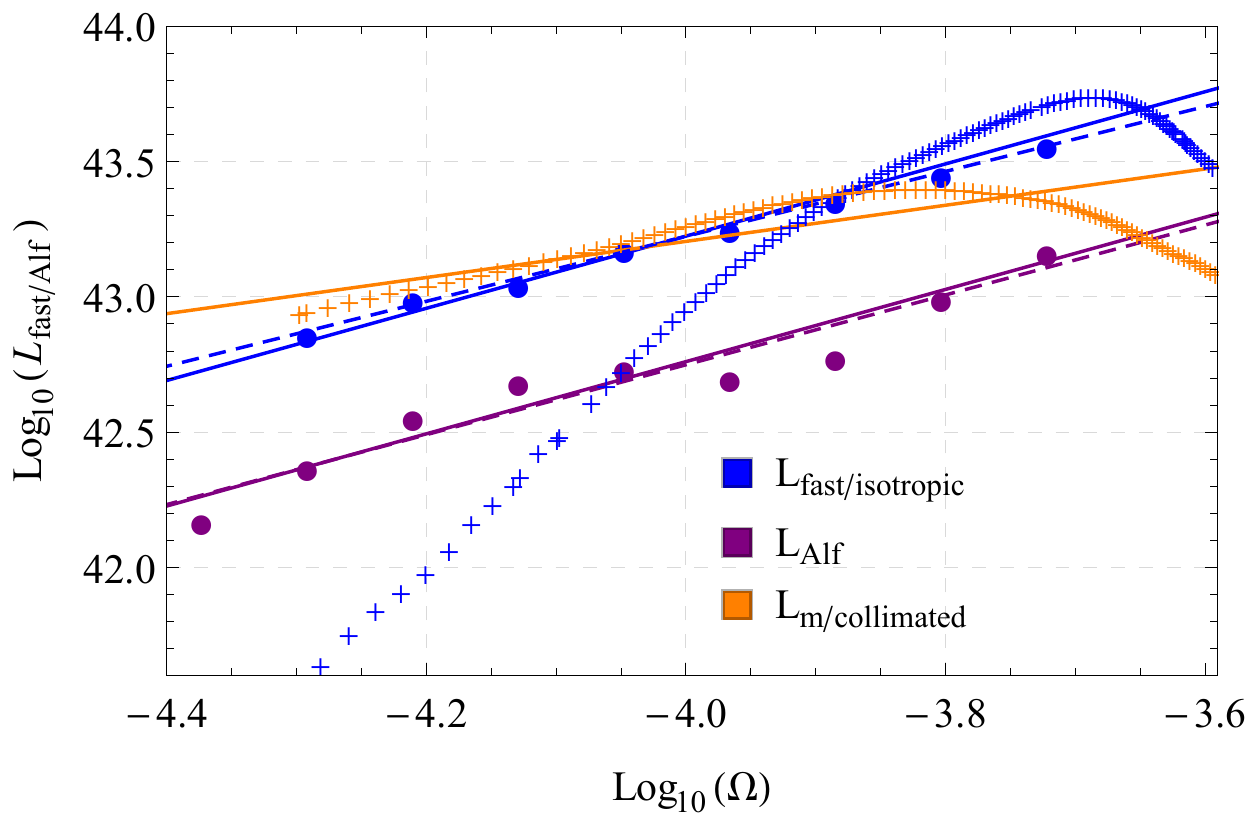}
\end{overpic}
\caption{The total fast-magnetosonic $\mathcal{L}_{\rm fast}$ (blue, in units of ergs/s), Alfv\'en wave $\mathcal{L}_{\rm Alf}$ (purple), and motion-driven (orange) luminosities as functions of the orbital frequency $\Omega$ (in units of 1/s). For $\mathcal{L}_{\rm fast}$ and $\mathcal{L}_{\rm Alf}$, the dots represent numerical integration results; the dashed lines are linear fits with flexible slopes; the solid lines are fits with a fixed slope of $4/3$. For $\mathcal{L}_{\rm m}$, the curve is from Eq.~\eqref{eq:MotionDriven} (although both terms are included, only the first term corresponding to collimated radiation is significant). For numerical data, the crosses are the measured isotropic and collimated fluxes taken from \cite{Neilsen:2010ax} for the non-spinning binary system. }
\label{fig:comparison}
\end{figure}

In Fig.~\ref{fig:comparison}, we plot the $\Omega$-dependent luminosities for both fast-magnetosonic and Alfv\'en waves, for an equal-mass binary system (as is simulated in \cite{Neilsen:2010ax}, \cite{Palenzuela:2010nf}, and \cite{Alic:2012df}), with the cut-off radius chosen at $1.5 $ times the horizon radius (it turns out that the results are insensitive to the cut-off radius).
More specifically, we substitute the density profiles appropriate for point masses following Newtonian Keplerian orbits into Eq.~\eqref{eqh}, and feed the resulting metric perturbation into the right hand side of Eq.~\eqref{eqwe} to obtain the expressions for $S_1$ and $S_2$. These then allow us to numerically integrate out Eq.~\eqref{eqreds} and acquire $\psi_1$ and $\psi_2$, representing the Alfv\'en and fast-magnetosonic waves respectively. To compute the Alfv\'en flux $\mathcal{L}_{\rm Alf}$, we apply $\partial_u$ and $\partial_v$ to $\psi_1$ and take the results through a numerically Fourier transformation procedure to obtain $\partial_u A^{\pm}$ and $\partial_v A^{\pm}$ according to Eq.~\eqref{eqfourier}. Finally, another numerical integration according to Eq.~\eqref{eqla} provides us with $\mathcal{L}_{\rm Alf}$. We do this for several black hole separations, as signified by their different Keplerian orbital frequencies, and plot the results as the purple dots in Fig.~\ref{fig:comparison}. We also compute the fast-magnetosonic fluxes $\mathcal{L}_{\rm fast}$ at these separations. In this case, we simply need to project $r\psi_2$ onto $\exp(im\phi)$ basis (taking $m$ up to $30$) and substitute the resulting $f_m$ values into Eq.~\ref{eqfm} to compute $\mathcal{L}_{\rm fast}$. The results are plotted as the blue dots in Fig.~\ref{fig:comparison}.

From the figure, we can see that the luminosity values are consistent with the quadrupolar contribution's dominance over higher multipoles, with a $\Omega^{4/3}$ scaling. We can also read off the dependence of $\mathcal{L}_{\rm Alf}$ and $\mathcal{L}_{\rm fast}$ on the magnetic field strength from their respective formula (Eqs.~\eqref{eqla} and \eqref{eqfm}), which is $B_4^2$. Simple dimensional consideration fixes the dependence on $M_8$ for us, which is $M_8^{10/3}$. What remains to obtain a formula similar to Eq.~\eqref{eq:MotionDriven} for the gravitation-driven case is the determination of the coefficients of proportionality, which set the overall amplitudes for the fluxes. These are simply the intercepts on the vertical axis of the solid purple and blue fitting lines in Fig.~\ref{fig:comparison} (in other words, they come from actually solving the equations and are not new independent rough estimates). In the end, we obtain that 
the gravitation-driven radiation should scale as
\begin{align} \label{eq:Fit}
\mathcal{L}_{\rm G} = &\mathcal{L}_{\rm fast} + \mathcal{L}_{\rm Alf} \notag\\
\approx  &1.7 L_{43} B^2_4 M^{10/3}_8 \Omega^{4/3}_{-4} 
+ 0.58 L_{43} B^2_4 M^{10/3}_8 \Omega^{4/3}_{-4}\,. 
\end{align}
Close to merger, the gravitation-driven, fast magnetosonic radiation dominates over flux contributions from Blandford-Znajek and kinetic-motion-driven radiations (Eqs.~\ref{eqr} and \ref{eq:MotionDriven}). This is consistent with the numerical observations of \cite{Neilsen:2010ax} and \cite{Moesta:2011bn} (see the top-right corner of Fig.~\ref{fig:comparison}). On the other hand, we caution that our computations do not take into account nonlinearities, so the analytical fit to numerical data should be interpreted with a pinch of salt. The aim of the present paper is only to demonstrate the existence of the gravitation-driven radiation, and the fact it can potentially produce large fluxes, especially an isotropic one during merger, rather than trying to make a fit to the numerical data with our zeroth-order calculation. In particular, our results should in no way be interpreted as fully ``explaining" the numerical results. 
In particular, we note that the matching for the fast-magnetosonic/collimated flux (blue crosses versus blue lines) at low frequencies is less accurate. Without a detailed examination involving targeted numerical experiments and higher order analytical computations, we can not state with certainty the exact reason for this, so future studies are required. Here, we can but point out some more obvious subtleties in the matching procedure.

Most importantly, as mentioned above, the Newtonian approximation breaks down in the vicinity of the black-holes in our zeroth-order calculation, and this happens regardless of the orbital separation. Although the fluxes change by only a few percentage points when we move the inner cut-off radius from $3\tilde{M}$ to $2\tilde{M}$, the dominant contribution to our numerical flux integrations nevertheless originate from the neighbourhoods of the black holes, instead of the wavezone. Therefore the omission of nonlinear relativistic effects might be the main approximation here, and taking into account the post-Newtonian or relativistic corrections may further change the luminosity estimates above. Other effects, such as the absorption by black holes, should also be treated properly. 

 Secondly, the numerical fluxes are divided according to their directions of propagation, catering more for the observational consequences than for matching with analytical classifications. Such imperfect correspondences between concepts employed by numerical and analytical studies 
lead to systematic matching errors. For example, the collimation in the numerical study is defined to be flux propagating inside a cone of a certain opening angle, in analogy with the usual jet language, while for Alfv\'en waves climbing the vertical magnetic field lines, a cylinder enclosing the binary (or two cylinders around individual black holes when they are far apart) would be more appropriate. Therefore, with a large extraction radius and when the black hole separation is large, the numerical cone would likely enclose a fair amount of fast-magnetosonic waves, contributing to the relative weakness of numerically measured isotropic flux. Many other numerical difficulties associated with subtracting off a background radiation in order to construct a division of the overall flux into the collimated and isotropic types, especially when the overall flux is weak, have been discussed in the numerical papers such as \cite{Neilsen:2010ax} and \cite{Moesta:2011bn}. We refer interested readers to these important literature. 

In the future, more specifically designed numerical experiments are necessary to test this gravitation-driven emission mechanism, including possibly binary star, instead of binary black hole, simulations. Improved sophistication in analytical computations is also necessary, before the effects of the various simplifying assumptions we made in the present work can be disentangled. 

\section{Discussion}
We briefly comment on plasma wave generation during the other stages of binary black hole coalescences. 
 During the merger phase, both the spacetime and the magnetosphere are highly dynamic, and the best tool to understand their evolution is through numerical simulations. However, in the ringdown stage, the time-dependent part of the emission arises from: (i) the ringdown of the magnetosphere, as described by its eigenwaves (\cite{Yang:2014zva,Yang:2015ata}); (ii) the gravitational quasinormal modes will drive additional emission by coupling to the stationary part of the black-hole jets, an effect quantifiable using black-hole perturbation theory.  Note that by the ``ringdown" stage, we mean the period before the post-merger black hole settles down to Kerr.  The settling time can be estimated as $1/\omega^I_{22}$, where $\omega^I_{22}$ is the imaginary part of the frequency for the $l=2, m=2$ quasinormal mode (the dominant mode). The value of $\omega^I_{22}$ is about $0.1/M$ for Schwarzschild black holes and $\omega^I_{22} \sim\sqrt{1-\bar{a}}/M$ for rapidly spinning black holes, which asymptotes to zero in the extremal spin limit (i.e. the modes are long lived and the settling is protracted) \footnote{For generic Kerr black holes, please see Fig. 5 in \cite{Yang:2012he} for the mode decay rates.}. 
For a post-merger black hole of $10^8$ solar masses, the Schwarzschild formula translates into a settling time of about eight and a half hours. So although extremely transient in nature, this period may be observationally detectable.  
On the other hand, the real part $\omega^R_{22}$ is $\sim 1/M$ for rapidly spinning black holes and $\sim 0.5/M$ for Schwarzschild black holes. During the ringdown stage, the gravitation-driven luminosity can be estimated as
\begin{align}
\mathcal{L}_{\rm G} \sim B^2 M^2 (\omega^R_{22} M)^{4/3} e^{-2 \omega^I_{22} t}\,,
\end{align}
while the Blandford-Znajek flux is approximately
\begin{align}
\mathcal{L}_{\rm BZ} \sim B^2 M^2 (a_{\rm f}/M)^2\,,
\end{align}
where $a_{\rm f}$ is the spin parameter for the final black hole. As the final black hole in generic binary mergers is rotating, we expect the Blandford-Znajek contribution to be important, and the gravitation-driven emission to also be an important part of the total flux at least within a timescale of $1/\omega^I_{22}$. 

During the ringdown stage, both the spacetime metric and the magnetosphere would be time-dependent, with similar but not exactly the same characteristic frequencies (\cite{Yang:2014zva}). The gravitation-driven mechanism would account for the metric variation's modifying effect to e.g., the Blandford-Znajek process, but not that from the magnetosphere ringing. In other words, multiple transient effects are present and it would be difficult to disentangle the signals they generate. Nevertheless, if quasi-periodic flux variations from the post-merger black hole can be detected, then one could in principal do interesting measurements such as that on the black hole spin.

For completeness, we can also estimate the flux modification due to the presence of current-sheets near the black holes, which is approximately the geometric mean of collimated and acceleration-induced radiations (see Eq.~42 in \cite{Brennan:2013ppa}). With units restored and according to Eq.~\eqref{eq:MotionDriven}, the corresponding luminosity is sub-dominant near merger. In addition, although we have examined the gravitation-driven plasma wave generation here in the context of force-free plasma, we expect similar signatures to persist in materials following more generic MHD equations.

Finally, we note that in the binary black hole example, energy is emitted at very low frequencies (below the plasma frequency). In fact, during the Blandford-Znajek process, the outgoing energy flux is carried out at the DC frequency. This is allowed for MHD waves (including waves in force-free plasma), but not for unmagnetized plasma (\cite{ThorneBlandford}).

\acknowledgments
We thank Luis Lehner and Ted Jacobson for discussions, reading over a draft of this manuscript and giving many useful comments. We also thank an anonymous referee for numerous important suggestions and comments. H.~Y.~acknowledges support from the Perimeter Institute for Theoretical Physics and the Institute for Quantum Computing. Research at Perimeter Institute is supported by the government of Canada and by the Province of Ontario through Ministry of Research and Innovation. F.~Z.~is supported by NSFC Grants 11443008 and 11503003, Fundamental Research Funds for the Central Universities Grant No.~2015KJJCB06, and a Returned Overseas Chinese Scholars Foundation grant. 

\bibliographystyle{apj}

\end{document}